# Polarised Photoluminescence from Surface-Passivated PbS Nanocrystals

M. J. Fernée, J. Warner, A. Watt, S. Cooper, N. R. Heckenberg, H. Rubinsztein-Dunlop

*Center for Quantum Computer Technology, Department of Physics, University of Queensland, Brisbane, Queensland 4072.*

**Abstract:**

Effective surface-passivation of PbS nanocrystals in aqueous colloidal solution has been achieved following treatment with CdS precursors. The resultant photoluminescent emission displays two distinct components, one originating from the absorption band-edge and the other from above the absorption band-edge. We show that both of these components are strongly polarised but display distinctly different behaviours. The polarisation arising from the band-edge shows little dependence on the excitation energy while the polarisation of the above-band-edge component is strongly dependent on the excitation energy. In addition, time resolved polarisation spectroscopy reveals that the above-band-edge polarisation is restricted to the first couple of nanoseconds, while the band-edge polarisation is nearly constant over hundreds of nanoseconds. We recognise an incompatibility between the two different polarisation behaviours, which enables us to identify two distinct types of surface-passivated PbS nanocrystal.

**Introduction:**

Zero dimensional semiconductors or quantum dots are promising candidates for a wide range of new technologies. There are many different techniques for achieving quantum dot behaviour, some of which have rather restricted operational regimes. Semiconductor nanocrystals (NCs) can make useful quantum dots that can operate in a wide range of environments and conditions. Robust high quality NCs with electrical and optical properties comparable to those achieved using self-assembly via molecular beam epitaxy can now be routinely synthesised using colloidal chemistry methods [1]. Furthermore, there is considerably more flexibility in the control of shape [2,3], size [4] and structure [5] using colloidal chemical synthesis, and considerably more scope for use in a wide range of applications. One very useful property that was first realised with highly controlled NC synthesis was the ability to tune the absorption band-edge and consequently the fluorescent emission wavelength via precise control of the nanocrystal size, which controls the confinement energy of the charge carriers [4]. This property is especially useful if the NCs are to be used as chromophores for labelling different biological molecules [1].

Of the various optical properties of NCs, polarisation has received comparatively little attention. This mainly stems from the fact that many NCs are spherical and for spherical NCs under most cases of excitation, the absorption dipole axis and the emission dipole axis are not related and so the random orientations of the NCs average the polarisation to zero. However, polarised emission from NCs is of considerable interest as both a probe of intrinsic charge carrier dynamics [6] and as a potential diagnostic for obtaining orientation information [7]. Recently it has been shown that only a slight change in the aspect ratio of a CdSe NC results in pure linearly polarised emission from a single NC at room temperature [7]. Polarised emission from bulk solutions is less commonly observed, although it has long been known that direct excitation of the lowest energy exciton state can produce polarised emission [6]. However, in general if the emission follows energy relaxation from higher excited states, any polarisation memory is usually lost. One exception is the polarisation memory exhibited in porous silicon [8-10]. In this case the polarisation could be traced directly to a combination of shape anisotropy and dielectric screening [9] and not to an intrinsic-state polarisation memory.

In this paper we show that colloidal solutions of surface-passivated PbS (SP-PbS) NC's can display strong photoluminescence (PL) that is partially polarised parallel to the excitation source polarisation. Two distinct types of polarisation behaviour are observed, corresponding to emission in two distinct spectral regions delineated by the NC absorption band-edge. We discuss our results in the context of known polarisation behaviour in other NC systems and use the different polarisation behaviours to identify two distinct types of SP-PbS.

**Surface Passivation.**

The relatively large surface to volume ratio of NCs can result in strong interactions with quantum confined charge carriers. This means the effect of the NC surface on the charge carriers cannot be neglected [11]. It is usual practice to modify the surface in some way in order to reduce or prevent the charge carriers interacting with the surface. Surface passivation with various organic ligands [6,12,13] or epitaxial overcoating with a wide band-gap semiconductor [11,14-17] can be used to enable efficient radiative recombination of the intrinsic charge carriers within the NC. Photoluminescence quantum efficiencies approaching unity have been achieved [1,13,15], indicating that non-radiative recombination pathways can be effectively suppressed using these strategies.

Lead sulfide (PbS) is a narrow band-gap semiconductor that can readily be prepared as NCs exhibiting strong quantum confinement of both charge-carriers [18,19]. For example, colloidal suspensions of PbS NC's can readily be prepared that exhibit a blue-shift of the absorption band-edge of nearly 1.5 eV, compared to the bulk [20,21]. Such a large blue-shift indicates a significant amount of confinement energy. This leads to the possibility of obtaining a size tunable band-edge that extends from the mid-infrared to the visible, covering the extremely important communications wavelengths.

Due to the strong quantum confinement of charge carriers, the smallest PbS NCs are most in need of effective surface passivation. As an example, there exists a simple aqueous colloidal synthesis that produces nearly spherical highly monodisperse PbS NCs of approximately 3 nm diameter [21]. These solutions exhibit a well known three peaked absorption spectrum, with the peaks attributed to excitonic resonances [19,22]. However, the PbS NCs so prepared display extremely weak PL, which has been attributed to poor surface passivation. There have been attempts at surface passivation using various organic ligands [12,23]. However strong surface interactions were still present, which strongly modified the absorption spectrum [23]. So far, the most successful technique for restricting surface interactions has been to grow larger NCs, which have a correspondingly smaller surface to volume ratio and a lesser degree of quantum confinement [24]. Otherwise, the poor fluorescence yield of small PbS NC's has restricted the study of strong quantum confinement to various nonlinear and absorption based techniques [22,23,25].

Surface passivation by epitaxial overcoating of a core NC with a large band-gap semiconductor has been developed for a number of NCs. For example, with CdSe NCs, epitaxial overcoating with either ZnS [14,15], CdS [16] or ZnSe [17] results in strong band-edge fluorescence with a pronounced improvement in the fluorescence quantum yield. In all cases, good epitaxial overcoating requires the use of a wide band-gap material with compatible lattice constants and crystal structure. This implicitly means using a material with the same crystal structure so that epitaxial overcoating can be accomplished in all crystal directions with the same efficiency.

The lead chalcogenides have a highly symmetrical rocksalt crystal structure, which is not common amongst most semiconductor materials and hence complicates the task of finding a compatible wide band-gap semiconductor suitable for epitaxial overcoating. We therefore investigated the possibility of using a material with a partially compatible crystal stucture. The zincblende structure, like the rocksalt structure, is a cubic structure based on a fcc arrangement of constituent atoms and differs only in the coordination of the nearest-neighbour atoms. Therefore, to a first approximation, any surface terminated by a single atomic species should be compatible with both the rocksalt and zincblende structures. For an ionic crystal, the compatibility should be complete, but for a covalently bound crystal, the bond rearrangement at the rocksalt/zincblende interface would be likely to complicate the interface. We chose CdS (zincblende 5.82 Å) for potential epitaxial overgrowth due to the small lattice mismatch with PbS (rocksalt 5.91 Å). In order to facilitate a good crystal match, our strategy was to promote the growth of PbS NCs with sulfur terminated surfaces using excess $S^{2-}$. We have used this technique to produce a range of surface passivated (SP-PbS) NCs with differing PL properties [26].

**Sample Preparation**

SP-PbS NCs are made according to the following procedure: Firstly aqueous colloidal suspensions of PbS NCs of approximately 3 nm diameter were prepared according to the method of Nenadovic and co-workers [21]. Surface-passivation of the PbS NCs was then accomplished by using a liquid phase epitaxial overcoating procedure that essentially adapts a well-known method for CdS NC synthesis as described by Spahnel and co-workers [27].

Typically, 100 ml of $2\times10^{-4}$ M solution of lead acetate (AR grade, Ajax) in 0.2% polyvinyl alcohol (22000 MW, BDH) was prepared using ultra-pure water (Millipore 18.2 M$\Omega$ cm). All chemicals were used directly without further purification. The solution was degassed by bubbling with Argon for 20 min, before the addition of 0.8 ml hydrogen sulfide gas. The amount of added $H_2S$ corresponds to more than a 2:1 molar excess and was used to promote the growth of PbS NCs with sulfur terminated surfaces. The solution was reacted for 5 minutes, forming a deep red PbS NC solution. The CdS overcoating procedure was then carried out as follows: Into the PbS NC solution, which has a pH of less than 5 in order to suppress individual CdS NC growth [5], is added 2 ml of a degassed 0.01 M sodium hexametaphosphate solution accompanied by rapid stirring. This was followed by the injection of a quantity of degassed $CdCl_2$ (AR grade, Ajax) solution sufficient to provide $Pb^{2+}$ to $Cd^{2+}$ ratio of 1:2. The injection of the $CdCl_2$ solution was either done slowly over a period of 2 minutes using a 0.01 M concentration solution or rapidly using a 0.1 M concentration. The solution was then reacted for 15 minutes with constant stirring. At this stage visible PL was detectable by eye when the solution was illuminated by a UV fluorescent lamp. A final PL activation process was usually performed [27] whereby the pH of the solution was raised above 10.5 by injecting 2 ml of a 0.1 M NaOH solution followed by the rapid injection of an excess 0.5ml of 0.1 M $CdCl_2$ solution. After this final stage a strong visible PL resulted.

The two variations in the addition of the $CdCl_2$ solution produce very different PL behaviors [26]. The slow addition of the more dilute solution tends to produce a deep red emission centred near 1.9 eV, while the rapid addition of the more concentrated solution results in broad PL emission that looks nearly white to the eye.

**Experimental methods**

Absorption spectra were obtained using a UV-Visible Spectrophotometer (Perkin-Elmer Lambda 40). Measurements were taken using a 10mm cuvette (both UV quartz and PMMA) with the samples diluted with ultra-pure water in order to ensure that absorption is kept within the Beer-Lambert limit. Appropriate solutions of PVA, sodium hexametaphosphate and PVA/hexametaphosphate mixtures were used for background subtraction in order to produce the final absorption profiles.

PL and photoluminescence excitation (PLE) spectroscopy were conducted using a commercial fluorimeter (Spex Fluoromax3). The slits for the excitation and detection channels were set to obtain a 3 nm bandpass. The PL spectra were corrected for the response of the photomultiplier tube. Time-resolved PL decays were obtained using a time-correlated single photon counting spectrometer (Picoquant Fluotime 200) with femtosecond pulses sourced from a frequency-doubled modelocked Ti:sapphire laser (Spectra Physics Tsunami). A Glan polariser was used to set the excitation polarisation and remove any residual fundamental from the excitation beam. A Glan polariser was inserted into the detection path for the polarisation-resolved measurements. The bandwidth of the detection path was set at 16 nm by using 2 mm slits with a dual grating subtractive monochromator.

The steady-state polarisation measurements were conducted using the fluorimeter with the addition of a Glan polariser in front of the excitation beam and a plate polariser analyser in front of the detection channel. This instrument detects the photoluminescence at 90 degrees to the direction of the excitation beam. The polarisation response of the detector channel was determined using the following procedure: The excitation polarisation was set parallel to the detection axis (horizontal with respect to the plane of the apparatus) after which the signal for two orthogonal analyser settings (horizontal and vertical with respect to the plane of the apparatus) were measured. In this geometry, symmetry dictates that the emission should not have any polarisation and so these two signals can be used to calibrate the polarisation-dependent response of the detection channel. This calibration technique was tested on a different system against an independent calibration using a halogen lamp. Both calibrations agreed to within ± 10%.

The polarisation response of the lifetime spectrometer was determined by a number of different methods depending on the sample being studied. For liquid samples, the polarisation response was determined absolutely using a solution of CdS NCs, which was pumped far above the absorption band-edge to ensure unpolarised emission. Under these circumstances, both polarisation-resolved lifetime curves were identical apart from a scaling factor that provided the calibration factor at the appropriate wavelength. This source was used to provide a series of absolute spot calibrations across the visible spectrum, which was necessary for solutions with long polarisation lifetimes. However, in the case of thin films observed at an acute angle, polarisation dependent reflections occur at the surface producing a detectable polarisation offset. Therefore, we used the technique of tail matching, which assumes that the emission is unpolarised after a sufficiently long time. While this technique is not as accurate as an absolute calibration, we found it reliable for much of the visible emission, where the tail of the decays matched almost exactly. However, for the low energy emission, we were unable to match the tail of the decays due to emergence of a long-time component in the polarisation. Therefore this data is not presented here.

**Results and discussion:**

The surface-passivation of PbS NCs involves a significant addition of $Cd^{2+}$ as well as excess $S^{2-}$ in order to promote the growth of a surface-protecting CdS coating. We monitored the effect of this process on the bare PbS NCs using absorption spectroscopy. The PbS NCs that are used as a starting material have a very distinctive three-peaked absorption spectrum [21]. These peaks have been attributed to absorption by discrete excitonic states [19,22]. In figure 1 we compare the absorption spectra obtained from pure PbS NCs, SP-PbS NCs and CdS NCs (made under conditions closely matching those used for the surface-passivation process, but without the presence of the PbS NCs). We note in figure 1 (a) that the SP-PbS NCs retain the three peaked spectrum which is attributed to PbS, but shows a slight blue-shift of the peaks as well as an increased absorption towards higher energies. Modifications to the absorption spectrum are also monitored by the second derivatives in figure 1 (b), which are more sensitive to peaks and edges. Here we see that the SP-PbS curve even more closely matches the pure PbS curve and shows no sign of any contribution by CdS NCs.

A slight red-shift of any excitonic features is expected when overcoating with a different material, due to an increase of the wavefunction envelope caused by penetration into the overcoat [11,14-16]. However, we observe a slight blue-shift of the peaks. In our case we suggest the blue-shift is due to the partial dissolution of the PbS surface by the $Cd^{2+}$ ions [28]. This effect is reasonably slow and is arrested by raising the pH of the solution during the activation phase of the surface-passivation process [26].

The surface-passivation produces two types of PL emission [26]. First we find emission from at and below the absorption band-edge near 2.1 eV. This is the simplest type of emission which conforms to the standard model of PL emission; that is absorption at or above the band-edge followed by relaxation to and emission from the band-edge. However, it is known that for most NCs, the band-edge emission is unpolarised if significant energy relaxation occurs before emission at the band-edge [7]. For instance, the emission obtained from spherical CdS NCs is unpolarised under high-energy excitation We have verified this property and in fact used it to calibrate the polarisation response of our lifetime spectrometer. We show in figure 2 that the band-edge emission obtained from SP-PbS is polarised over a wide range of excitation energies. In figure 2 (a) we show the steady-state polarisation properties along with the polarised PL emission curves and an associated PLE curve (not polarisation resolved). Here the degree of polarisation, $\rho$, as defined is,

$\rho = (I_\parallel - I_\perp)/(I_\parallel + I_\perp)$,

where $I_\parallel$ and $I_\perp$ represent the signals obtained with the polariser orientated parallel and perpendicular to the polarisation of the excitation source. The polarisation of the PL spectrum is reasonably constant, only slowly reducing towards lower energies, where we expect trap-state emission to dominate. The polarisation of the PLE spectrum is also reasonably flat over a wide range of excitation energies, indicating very little dependence on the excited state. In figure 2 (b) we show the time resolved PL decays for the two orthogonal analyser settings, along with the polarisation associated with these curves. Here we see that while the PL decay curves are highly multi-exponential, they nevertheless maintain a nearly constant polarisation over many hundreds of nanoseconds. The slight fall in polarisation over the time period covered is attributable to the gradual reorientation of the NCs in the solution [29]. The polarisation that we observe for the band-edge emission is mostly independent of both excitation energy and emission processes. Such behaviour can be described in terms of a shape anisotropy/dielectric screening model [9,10]. In general, the PbS NCs that we make are slightly ellipsoidal with a long axis to short axis ratio of 1.2, as determined by an electron microscopy analysis. Accordingly, the dielectric mismatch between the PbS core and the outer passivating layer can give rise to a strongly shape dependent emission profile. Such an effect has been used to explain the origin of polarised emission in porous silicon [9,10].

A second type of emission can also be obtained with SP-PbS NC; that is strong emission from above the SP-PbS NC band-edge. An example of such above-band-edge emission is displayed in figure 3 (a). In fact such solutions often display both band-edge and above-band-edge emission features, as shown here. The above-band-edge component of the emission is characterised by a strongly peaked excitation spectrum that shows little contribution below 2.8 eV. The peak of this excitation spectrum near 3.3 eV also correlates very well with the second excitonic peak in the absorption spectrum as well as a similar feature in the PLE spectrum in figure 2 (a). Accordingly we attribute the above-band-edge emission to absorption by SP-PbS NCs [26]. In figure 3 (b) we show the degree of polarisation as a function of emission energy obtained after excitation with a range of energies. The non-polarised PL spectra are also included for comparison. Here we clearly see two regions displaying different polarisation behaviour. The polarisation of the above band-edge region is both strongly emission energy dependent as well as excitation energy dependent. The excitation dependence is explored in greater detail in figure 3 (c). Here we see a clear correlation between the polarisation of the above-band-edge emission (monitored at 2.33 eV) and the 3.5 eV peak in the excitation spectrum. The polarisation of the band-edge emission at 2.0 eV displays what seems to be a composite of the above-band-edge behaviour and the band-edge behaviour (as shown in figure 2), as both a strong excitation energy dependence as well as a region mostly independent on the excitation energy are evident.

Further investigations of the above-band-edge polarisation characteristics were undertaken using polarisation resolved PL lifetime spectroscopy. In this case the SP-PbS NCs were embedded in a thick polyvinyl alcohol (PVA) film, so that they could be mounted and cooled to 77 K in a liquid nitrogen cryostat, substantially increasing the PL output. The PL emission obtained from this sample is shown in the inset to figure 4 (b) and the points along this emission spectrum that were used for the time resolved studies are indicated appropriately. In figure 4 (a) we show the PL decays obtained and also include for comparison, a dotted line indicating a decay time-constant of 1.5 ns. It is apparent that the decays are strongly multi-exponential. At least 4 exponentials are required in order to obtain a reasonable fit. However, in general a more sophisticated multi-exponential lifetime analysis is required to determine a lifetime distribution [30]. Such an analysis is not necessary here, as we clearly see from these curves that the short time behaviour is dominated by a 1.5 ns decay time constant, while the lower energy emission shows an increasing long lifetime component. In figure 4 (b) we show the polarisation anisotropy associated with each decay curve in figure 4 (a). Here we see that the polarised emission is associated with the short lifetime components, as curves showing increased long lifetime decay components have progressively shorter periods of strong polarisation. Therefore, in contrast to what was observed for the band-edge emission, the above-band-edge polarisation is strongly time-dependent. As a rough guide, the departure of the decay curves from the 1.5 ns reference decay corresponds reasonably well to the respective width of the polarisation curves. This suggests that the polarised components have lifetimes in the nanosecond regime, while the long lifetime components are not polarised. These lifetime results are also consistent with the emission energy dependence of the steady-state polarisation shown in figure 3 (b), as the relative increase of the long lifetime component with decreasing energy will lower the steady-state polarisation accordingly.

In general the polarisation behaviour of the above-band-edge emission has the character of emission following excitation at an absorption band-edge, where direct excitation of the emitting state is accomplished. However, in our case excitation is followed by significant energy relaxation before emission. We also note that the 1.5 ns decay time is characteristic of either excitonic or shallow trap lifetimes, while the unpolarised microsecond components are more likely to correspond to deep trap emission. Therefore the polarisation results indicate that the observed above-band-edge emission is not related to deep trap states, but rather some other mechanism. We note that the second excitonic peak in the PbS NC spectrum has been attributed to a $P_e$-$P_h$ transition [19]. If this is so, then the asymmetric wavefunctions could promote asymmetric trapping of one of the charge-carriers. Such charge trapping may involve either a localised shallow state or trapping within the passivating layer.

Finally we note that the two distinct types of polarised emission that we observe with SP-PbS NCs are incompatible. The polarisation caused by the combination of dielectric screening and shape anisotropy is essentially a static effect and should be present irrespective of the charge-carrier dynamics and other processes. However, we have shown that the polarisation of the above-band-edge emission shows no indication of such static effects. Therefore, we are led to suggest that different types of SP-PbS NCs are responsible for these two types of polarisation behaviour. Clearly, the ability to produce SP-PbS NC that only produce red band-edge emission indicates that the onset of above-band-edge emission corresponds to the emergence of a different type of NC. We have shown previously [26] that the above-band-edge emission can be attributed to SP-PbS NC and not CdS NCs. In fact, whenever CdS NC have been formed in the surface passivation process, they have been easily detected using both absorption spectroscopy and PL spectroscopy. Any further discussion on the nature of the structural differences in SP-PbS NCs that produce the two types of behaviour we have reported must await the results of structural studies that we are currently conducting.

**Conclusion.**

We have shown that SP-PbS NCs can exhibit strongly polarised PL emission. Two distinct types of polarised emission were observed which were associated with different spectral regions. The polarisation of the band-edge emission showed negligible dependence on the excitation energy and the polarisation remained nearly constant over the entire PL lifetime (even though the PL decay curves were strongly multi-exponential). We have therefore attributed the polarisation of the band-edge to a combination of shape anisotropy and dielectric screening effects. The polarisation of the above-band-edge emission is dramatically different from that observed for the band-edge. The polarisation shows a strong dependence on both the excitation energy and emission energy. Photoluminescence lifetime measurements revealed a polarised 1.5 ns lifetime component and an unpolarised microsecond lifetime component, with an increasing contribution from microsecond component towards decreasing emission energy. The strong excitation energy dependence and lifetime dependence of the polarisation of the above-band-edge emission is incompatible with the dielectric screening and shape anisotropy mechanism. Furthermore, significant energy relaxation prior to emission suggests that an intrinsic excitonic polarisation memory is unlikely. Therefore we have proposed that the polarisation of the above-band-edge component of the SP-PbS NC PL spectrum may be the result of rapid asymmetric charge trapping of one of the charge carriers following excitation of a $P_e$-$P_h$ excitonic state. Finally, the observation of two distinct and mutually incompatible polarisation behaviours indicates that there must be two types of SP-PbS NC. Structural investigations are currently underway in order to determine the precise differences between the two types of SP-PbS.

**Acknowledgements:** This work was supported by the Australian Research Council. We thank Dr Jamie Riches for his TEM studies.

**References:**
1. X. Michalet, et al., *Single Mol.* **2**(4), 261 (2001).
2. V. F. Puntes, K. M. Krishnan, A. P. Alivisatos, Science **291**, 2115 (2001); L. Manna, E. C. Scher, A. P. Alivisatos, J. Am. Chem. Soc. **122**, 12700 (2000).
3. S.-M. Lee, Y. Jun, S.-N. Cho, J. Cheon, J. Am. Chem. Soc. **124**, 11244 (2002).

4. A. P. Alivisatos, J. Phys. Chem. **100**, 13226 (1996).
5. A. Mews, A. V. Kadavanich, U. Banin, A. P. Alivisatos, Phys. Rev. B **53**(20), R13242 (1996).
6. M. G. Bawendi, P. J. Carroll, W. L. Wilson, L. E. Brus, J. Chem. Phys. **96**(2), 946 (1992).
7. J. Hu, *et al.*, *Science* **292**, 2060 (2001).
8. S. V. Gaponenko, *et al.*, Appl. Phys. Lett. **67**(20), 3019 (1995).
9. D. Kovalev, *et al.*, *Appl. Phys. Lett.* **67**(11), 1585, (1995).
10. J. Diener, Y. R. Shen, D. I. Kovalev, G. Polisski, F. Koch, *Phys. Rev. B* **58**(19), 12629 (1998).
11. Y. Cao, U. Banin, Angew. Chem. Int. Ed. **38**(24), 3692 (1999).
12. A. A. Patel, *et al.*, J. Phys. Chem. B **104**, 11598 (2000).
13. B. L. Wehrenberg, C. Wang, P. Guyot-Sionnest, J. Phys. Chem B **106**, 10634 (2002). ; H. Du, *et al.*, Nano Lett. **2**(11), 1321 (2002).
14. M. A. Hines, P. Guyot-Sionnest, J. Phys. Chem. **100**, 468 (1996).
15. B. O. Dabbousi, *et al.*, J. Phys. Chem. B **101**, 9463 (1997).
16. X. Peng, M. C. Schlamp, A. V. Kadavanich, A. P. Alivisatos, J. Am. Chem. Soc. **119**, 7019 (1997).
17. P. Reiss, J. Bleuse, A. Pron, Nano Lett. **2**(7), 781 (2002).
18. F. W. Wise, Acc. Chem. Res. **33**, 773 (2000).
19. I. Kang, F. W. Wise, J. Opt. Soc. Am B **14**(7), 1632 (1997).
20. S. Gallardo, M. Gutierrez, A. Henglein, E. Janata, Ber. Bun. Phys. Chem. **93**, 1080 (1989).
21. M. T. Nenadovic, M. I. Comor, V. Vasic, O. I. Micic, J. Phys. Chem. **94**, 6390 (1990).
22. J. L. Machol, F. W. Wise, R. C. Patel, D. B. Tanner, Phys. Rev. B **48**(4), 2819 (1993).
23. L. Guo, *et al.*, *Opt. Mat.* **14**, 247 (2000).
24. Wundke K., *et al.*, Appl. Phys. Lett. **75**(20), 3060 (1999).
25. Olkhovets A., Hsu R. C., Lipovskii A., Wise F. W., Phys. Rev. Lett. **81**(16), 3539 (1998); Krauss T.D., Wise F. W., Tanner D. B., Phys. Rev. Lett. **76**, 1376 (1996);

Krauss T.D., Wise F. W., Phys. Rev. Lett. **79**, 5102 (1997); Krauss T.D., Wise F. W., Phys. Rev. B **55**, 9860 (1997).

26. M. J. Fernée, A. Watt, J. Warner, S. Cooper, N. Heckenberg, H. Rubinsztein-Dunlop, cond-mat/0303267.
27. L. Spahnel, M. Haase, H. Weller, A. Henglein, J. Am. Chem. Soc. **109**, 5649 (1987); A. Eychmuller, A. Hasselbarth, L. Katsikas, H. Weller, J. Lumin. **48/49**, 745 (1991).
28. Moriguchi I., *et al.*, J. Chem. Soc., Faraday Trans. 94(15), 2199 (1998).
29. J. R. Lakowicz, I. Gryczynski, Z. Gryczynski, K. Nowaczyk, C. Murphy, Anal. Biochem. **280**, 128 (2000).
30. A. M. Kapitonov, *et al.*, J. Phys. Chem B **103**, 10109 (1999).

Figure 1. Absorption comparison of (a) pure PbS NCs (black), SP-PbS NCs (blue) and CdS NCs (red) made under conditions similar to those used for surface passivation. (b) the second derivative of the absorption curves shown above.

Figure 2. Polarisation of the band-edge emission. (a) The upper plot shows both the polarisation resolved PL data (red) resulting from excitation at 3.1 eV and the PLE spectrum (blue) corresponding to the emission at 1.94 eV. The emission polarisation is measured both parallel and perpendicular to the polarisation of the excitation source. The lower plot shows the degree of polarisation associated with the above PL (red) and PLE (blue) curves. (b) Time resolved PL for emission polarised parallel (blue) and perpendicular (red) to the pump polarisation, obtained from a solution at 300 K following excitation at 3.26 eV. The degree of polarisation determined from the two polarisation components is also shown (circles).

Figure 3. Polarisation of the above-band-edge emission. (a) PL (red) and PLE (blue) curves associated with the SP-PbS solution used in the polarisation studies. Arrows indicate excitation energies for PL and emission energies used for the PLE curves. (b) The degree of polarisation as a function of emission energy corresponding to excitation at

3.44 eV (triangles), 3.26 eV (squares) and 3.10 eV (circles). The corresponding PL curves are also shown for comparison. (c) The degree of polarisation as a function of excitation energy corresponding to emission at 2.33 eV (circles) and 2.00 eV (triangles). The corresponding PLE curves are also shown for comparison.

Figure 4. Time resolved polarisation. (a) The time resolved PL decays following excitation at 3.18 eV. SP-PbS NCs were embedded in a PVA polymer film and cooled to 77 K in a liquid nitrogen cryostat. The emission was monitored at 2.95 eV, 2.76 eV, 2.58 eV, 2.48 eV, 2.34 eV, 2.25 eV, and 2.14 eV. A decay curve with a lifetime of 1.5 ns is also included for comparison (dotted line) (b) The time resolved degree of polarisation corresponding to the emission that was monitored. (Inset. PL curve obtained from the sample with excitation at 3.18 eV. The points along the PL curve used for time resolved measurements are indicated with coloured circles.)

**Figure 1**

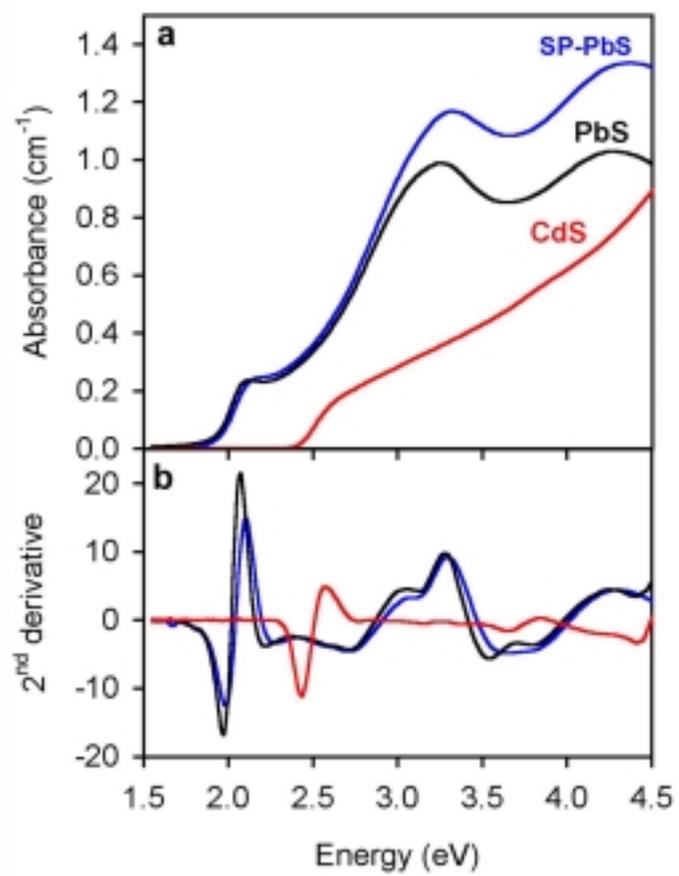

Figure 2

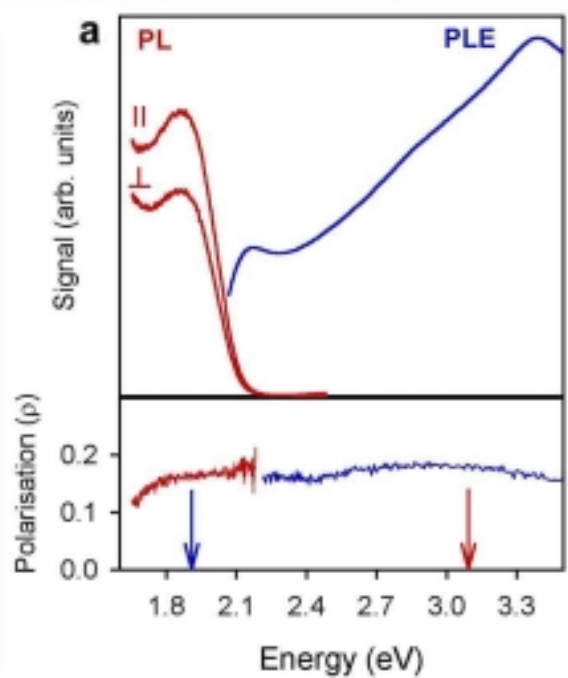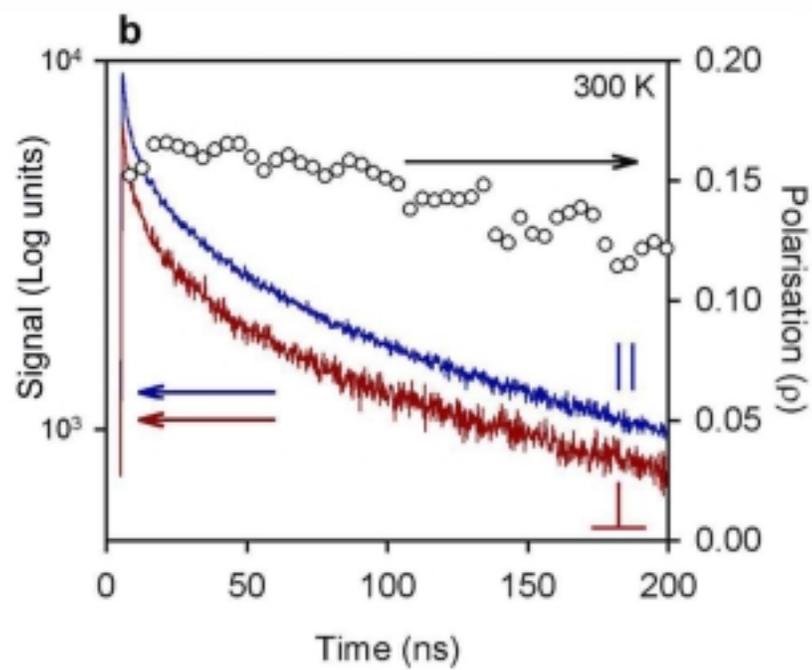

# Figure 3

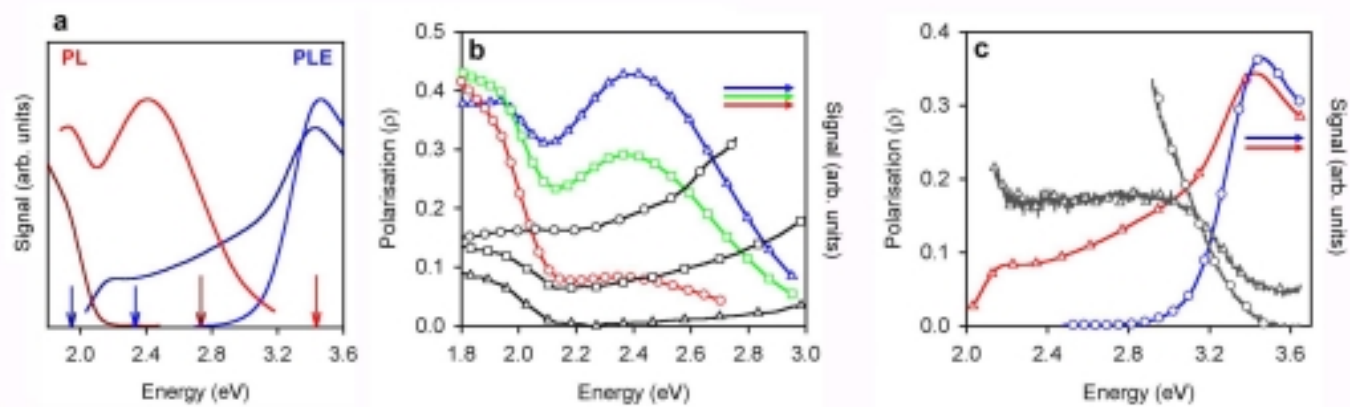

# Figure 4

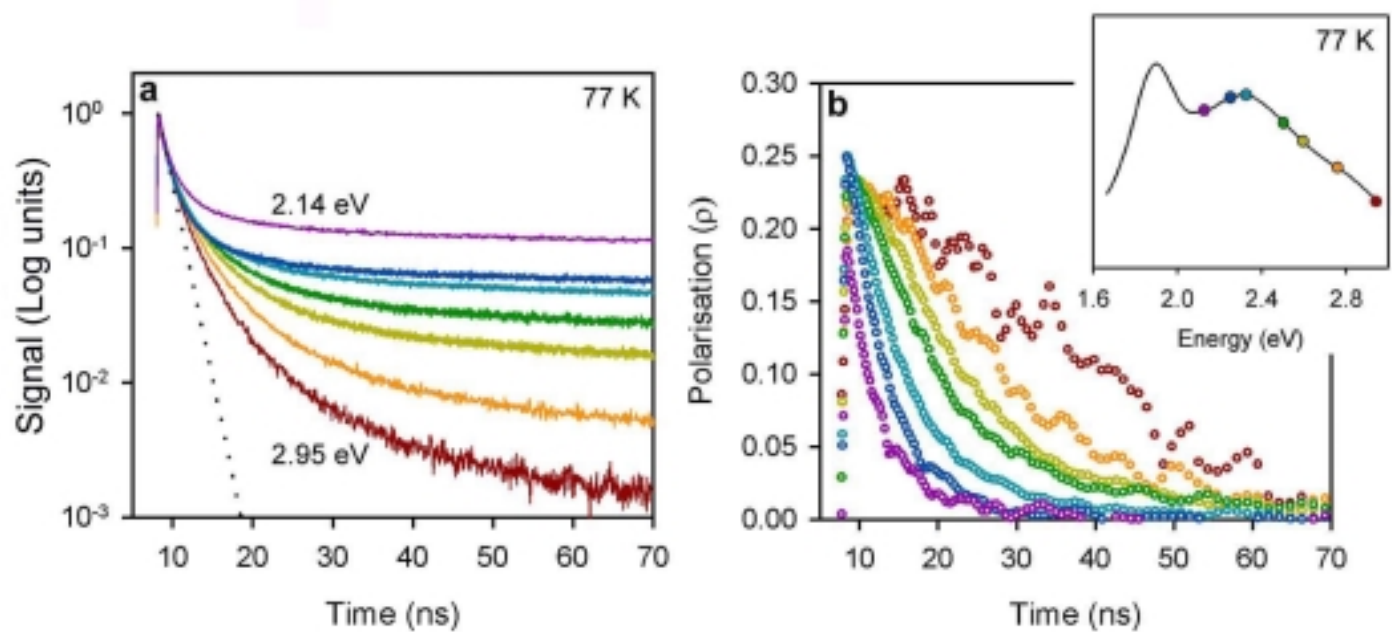